\documentclass[11pt,english]{article}

\usepackage[margin=1in]{geometry}

\usepackage[utf8]{inputenc} 
\usepackage[T1]{fontenc}
\usepackage{graphicx}
\usepackage{enumitem}
\usepackage{xcolor}
\definecolor{DarkBlue}{rgb}{0.1,0.1,0.5}
\definecolor{DarkGreen}{rgb}{0.1,0.5,0.1}
\usepackage{hyperref}
\hypersetup{
   colorlinks   = true,
 linkcolor    = DarkBlue, 
 urlcolor     = DarkBlue, 
	 citecolor    = DarkGreen 
}

\usepackage[capitalize]{cleveref}
\usepackage{booktabs}
\usepackage{subcaption}

\usepackage{cite}

\begin{document}
\title{\textbf{The Pedagogy of AI Mistakes:\\ Fostering Higher-Order Thinking}}
%
%
\author{Hadi Hosseini\\
Pennsylvania State University\\ 
University Park, USA\\
\texttt{\url{hadi@psu.edu}}
}
\date{}
\maketitle              
\begin{abstract}
As generative AI becomes increasingly integrated into higher education, its frequent errors and hallucinations, often seen as limitations, offer a unique pedagogical opportunity. By framing AI as a ``learning companion'' whose imperfect outputs prompt analysis, evaluation, and reflection, we argue that instructors can engage students in the fundamental processes of higher-order thinking. This paper presents a design-oriented study in which an AI-integrated syllabus in a \textit{database design} course deliberately leverages AI's limitations to foster critical thinking and higher-order cognitive skills aligned with Bloom's taxonomy of learning. Using a mixed-methods approach, we examine how structured interaction with AI-generated errors supports metacognitive engagement, reinforces disciplinary rigor, and relates to students' perceived AI literacy and subject-matter competency.

\end{abstract}

\section{Introduction}

Generative Artificial Intelligence (AI), and in particular Large Language Models (LLMs), has sparked significant debate about its role in education.
Recent works demonstrate both the growing presence of AI tools in learning contexts and their effectiveness in supporting student task completion \cite{cambaz2024use,prather2023robots,finnie2022robots}. In computer science education, LLMs are increasingly introduced from the first day of instruction in introductory programming courses \cite{porter2024learn} and are being explored across a range of paradigms, including collaborative and pair-programming settings \cite{denny2023chat,kazemitabaar2023studying}. 
At the same time, prior studies caution that AI-generated code can obscure underlying concepts, as students may successfully produce solutions they do not fully understand \cite{kazemitabaar2023studying}. These developments have amplified debates within the education community, with some advocating for restriction or prohibition of student AI use \cite{lau2023ban,denny2024computing}. However, such approaches risk overlooking both the inevitability of AI’s integration into educational practice and its potential to serve as a pedagogical companion when thoughtfully designed.

While over-reliance on AI can impede the development of core cognitive skills \cite{qian2025pedagogical}, its scaffolded use can position AI tools as ``learning companions'' that support deeper conceptual understanding and critical thinking \cite{loksa2022metacognition,prather2024widening}.
Because LLMs can produce errors, hallucinations, or context-insensitive responses, effective instruction must extend beyond domain knowledge to include the ability to evaluate, contextualize, and revise AI-generated outputs.

Guided by Bloom's taxonomy \cite{krathwohl1964taxonomy}, we argue for a shift in syllabus design and classroom practice, from the foundational cognitive levels of knowledge, comprehension, and application toward the higher‑order processes of analysis, evaluation, and synthesis.\footnote{Other frameworks such as Webb's Depth of Knowledge \cite{webb1997research} and Marzano's taxonomy \cite{marzano2007taxonomy} also encompass higher-order cognitive skills, including analysis and evaluation.} 
In this sense, \textbf{the fallibility of LLMs becomes pedagogically productive}: by requiring students to interrogate and validate AI-generated responses, instructional activities can actively engage learners in higher-order reasoning and critical judgment, aligning AI use with established learning theory.
Students learn more effectively by actively critiquing and revising solutions than by passively receiving feedback \cite{vanlehn2011relative, hosseini2019you, hosseini2019learning}. Thoughtfully integrated AI, particularly through its imperfect or error-prone outputs, can further foster reflective engagement with course content, supporting deeper mastery.

In this course-redesign study, we integrate AI-centric activities such as critical evaluation of model outputs, AI-guided problem decomposition, and structured reflection on AI use within core instructional tasks. 
We ask the following research questions: 
\begin{enumerate} [label=(\roman*)]
    \item {How can generative AI be meaningfully integrated into syllabus design to promote higher-order thinking, as guided by Bloom’s taxonomy?}
    \item {What role do students’ prior AI and domain knowledge (self-reported and actual) play in shaping their engagement with AI-integrated activities?}
\end{enumerate}

\subsection{Contributions}
We anchor our investigation in the context of an undergraduate database design course, using it as a concrete framework to explore AI‑integrated pedagogy. We introduce a course design that integrates generative AI activities, e.g., prompt engineering, error‑analysis exercises, and themed case studies, into each module to foster analysis, evaluation, and creation. 
In particular, we make the following contributions: 
\begin{enumerate}
\item 
We present a redesign of an undergraduate database course that integrates generative AI as a pedagogical companion to explicitly promote \textbf{metacognitive} skills, including self-monitoring, error detection, and critical evaluation of AI-generated solutions, thereby shifting learning emphasis toward \textbf{higher-order cognitive} processes.
\item 
We propose a mixed-methods evaluation framework that combines \textbf{self-reported literacy}, objective \textbf{AI competency} measures, and subject-matter performance assessments, enabling a nuanced examination of how students engage with and learn alongside generative AI.
\item
Through pre/post assessments and correlational analyses, we demonstrate gains in database competency and identify \textbf{systematic misalignment} between students' perceived AI abilities and their objectively measured performance, highlighting the pedagogical importance of calibration and reflective judgment in AI-integrated learning.
\end{enumerate}

\section{Course Structure: an AI-Integrated Design}

Our design is guided by the premise that higher-order cognition is best developed through structured opportunities for evaluation and revision, particularly when learners engage with imperfect responses.

\subsection{Core Topics}
The course is structured into a series of several interconnected modules that span one or two weeks. 
Topics include: introduction to Database Management System (DBMS) architecture, Entity–Relationship (ER) modeling, schema normalization, relational databases and SQL, NoSQL, key‑value stores, and graph databases, culminating in a capstone integration project.

\subsection{Module Components}
There are four core components that support scaffolded learning---lesson plans, short videos, AI‑focused exercises, and curated readings.
These components form a cohesive design that systematically advances students from foundational knowledge to higher‑order analysis, evaluation, and creation, aligning with Bloom's Taxonomy of learning.

\subsection{Lesson Plans: What, Why, and How}
Each module begins with a detailed lesson plan that states explicit \textbf{learning outcomes}, clarifying what students should be able to do (e.g., ``Design and implement a 3NF schema for a real‑world scenario'').  We frame these outcomes through three guiding questions:
\begin{enumerate}
\item
\textbf{What?} the core concepts (e.g., ``What is second‑normal form?''), 
\item 
\textbf{Why?} the pedagogical rationale (e.g., ``Why does normalization reduce data redundancy and improve integrity?''), and 
 \item 
\textbf{How?} the procedural mechanics (e.g., ``How do you decompose relations to eliminate partial dependencies?'').
\end{enumerate}
This What/Why/How structure emphasizes metacognitive reflection and primes students to connect theory with practice.

\subsection{Short Video Components}
Each module includes a 4–10 minute introductory video that motivates the topic and sketches the big picture. These are supplemented by one or two supporting video clips; for example, a screencast demonstrating interactive normalization of a sample ER diagram or a short interview with industry practitioners on graph query use cases. 
Students view these videos before tackling hands‑on exercises, ensuring a flipped‑classroom dynamic.

\subsection{AI Modules}
Central to our design is a weekly {AI Module}, structured to draw on Bloom's taxonomy and metacognitive theory through an iterative critique-refinement cycle. 

\begin{itemize}
\item
\textbf{Prompting Strategies}: Students engage at the level of \emph{application} by learning techniques such as zero-shot vs. few-shot prompting, chain-of-thought prompting, and temperature tuning \cite{dong2022survey}. 
 For example, when generating SQL snippets, students learn to include schema definitions and sample rows in their prompt to elicit accurate SELECT statements.
 \item 
\textbf{Failure-Mode Analysis}: To promote \emph{analysis} and \emph{evaluation}, we present ``what-can-go-wrong'' cases such as an AI‑generated ER diagram that violates referential integrity or produces nonsensical attribute names, and guide students in diagnosing and critiquing these outputs.
\item 
\textbf{Case Study} (Santa's Workshop): Students integrate these skills in a recurring scenario, iteratively \emph{creating}, evaluating, and refining database designs. They craft prompts to generate schema suggestions, normalize tables, or write queries, then iteratively refine their prompts based on AI feedback and conducting error analysis.
\end{itemize}

Across these components, AI's imperfect outputs serve as catalysts for evaluation, positioning critique as the central mechanism driving higher-order cognition. 
The resulting cycle of generation, evaluation, and revision fosters deeper conceptual understanding and metacognitive engagement.

\subsection{Reading Modules}
Complementing the interactive and video materials, each module includes a curated {Reading Module}, comprising mandatory and optional resources such as textbook excerpts, online articles, or concise tutorials. 
For example, the normalization week directs students to a classic 1970 paper on relational design, while the graph week offers a practical Neo4j tutorial. 
These readings reinforce the foundations and provide deeper context for exercises.

\subsection{Assessment}
Both \textit{formative} and \textit{summative} assessments are employed to provide frequent feedback, e.g. weekly homework and quizzes covering both database concepts and AI components, as well as cumulative midterm and final exams.

\section{Study Design and Findings}
Students ($n = 13$) were enrolled in an asynchronously delivered undergraduate database design course emphasizing self-directed learning.\footnote{Self-directed learning is a prevalent approach in online education; it presents greater challenges---especially with the rise of LLMs---compared to traditional lecture-based instruction, as it is more susceptible to the ``illusion of learning'' \cite{bao2023mind}.}
They completed a \textbf{prior knowledge survey} combining self-reported AI and database literacy (5-point Likert) with objective multiple-choice questions on both concepts. 
%
%
This approach captured initial confidence versus actual knowledge, measures learning gains, and examines the link between AI literacy and subject-matter mastery.
The experimental details and the full set of assessment questions are provided in \Cref{sec:literacy} and \Cref{app:subjectmatter}.

\begin{figure}[t]
\begin{subfigure}{0.45\textwidth}
    \centering    \includegraphics[width=.9\linewidth]{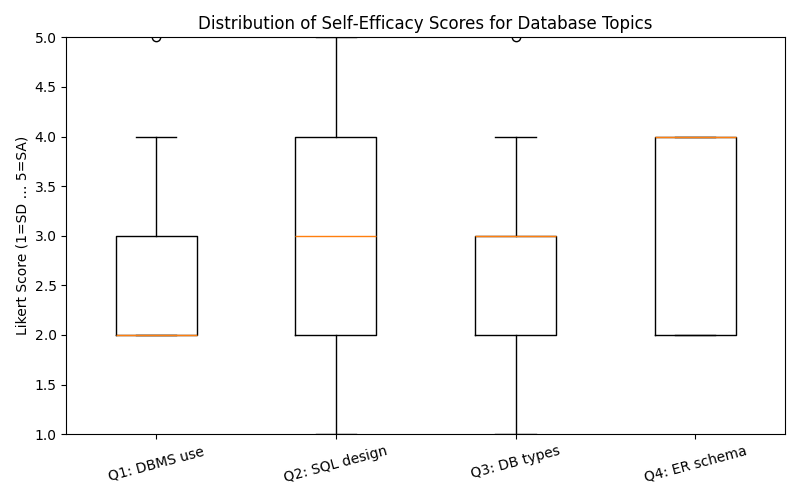}
    \caption{Self‑Efficacy for Database Literacy}
    \label{fig:DBLit}
\end{subfigure}
\hfill
\begin{subfigure}{0.45\textwidth}
    \centering
\includegraphics[width=.86\linewidth]{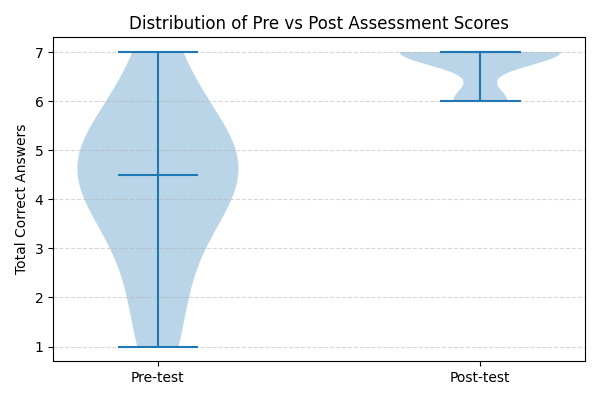}
    \caption{Pre/Post Assessment Scores}
    \label{fig:prepost}
\end{subfigure}
\caption{Distributions of (a) self-efficacy scores and (b) pre/post assessment scores.}
\end{figure}

\subsection{AI Literacy and Competency}
We measured AI literacy/competency with a two-part questionnaire: (i) six  self-assessment questions capturing \textbf{self-reported AI literacy} by adapting questions from the AILQ (Artificial Intelligence Literacy Questionnaire) \cite{ng2024design} and MAILS (Measuring AI Literacy and Self-efficacy) \cite{carolus2023mails} questionnaires, and (ii) five objective multiple-choice questions covering three core generative AI domains---conceptual knowledge (e.g.  `tokens' and `prompts'), prompt engineering (e.g. few‑shot prompting and format instruction), and model adaptation (e.g. fine‑tuning concepts)---yielding an independent \textbf{AI competency score} that reflects demonstrated knowledge.

We analyzed student responses across domains. Agreement is strongest in \textit{ethics} (67\% agreement, 0\% disagreement) and \textit{evaluation} (66\%, 8\%), moderate in \textit{application} (50\%, low disagreement), and weakest in \textit{self-efficacy} (45.5\%, 4\%), indicating uncertainty in applying AI to complex tasks and motivating more scaffolded, hands-on and critique-based learning.
Additionally, correlating self-reported AI literacy with objective competency, we find a moderate negative correlation ($r=-0.39$) that is not statistically significant ($p\approx 0.21$), but could suggest potential overestimation of AI abilities among students.

\subsection{DBMS Self-Reported Literacy}\label{sec:subjectmatter}

Figure \ref{fig:DBLit} reveals distinct confidence profiles across the four database topics. 
Students report the lowest and most tightly clustered self‑efficacy in \textbf{DBMS use} (median=2), indicating a general lack of confidence with database software. \textbf{SQL design} shows the greatest variability (scores spanning 1–5 and a wide interquartile range), suggesting uneven experience with query construction. Familiarity with \textbf{database types} centers around neutrality (median=3) with moderate dispersion, reflecting mixed comfort across relational, document, and graph systems. Finally, \textbf{ER schema} design emerges as the area of strongest consensus (median=4), though a notable lower whisker indicates a subset of learners still uncertain. These patterns pinpoint where targeted AI‑supported scaffolds—especially for DBMS fundamentals and SQL practice—could most effectively bolster student confidence.

\subsection{Pre/Post Assessment}

We evaluated learning gains using a \textbf{pre/post assessment} with seven identical multiple-choice questions administered at the start of the course and later embedded in the midterm and final. This design enables direct comparison of performance before and after instruction. The items cover core database concepts, including ER modeling, normalization, schema design, and basic queries.
On the pre‑test (out of 7), scores ranged broadly from 1 to 7 (mean = $4.25$, SD=$1.57$), with a median of 4 and scores distributed across the full range. In contrast, the post‑test distribution is tightly clustered between 6 and 7 (mean = $6.83$, SD $\approx0.39$), see Fig. \ref{fig:prepost}. A paired t‑test confirms this improvement is highly significant ($t(12) \approx5.10, p<.001$), with a mean score increase of 2.58 points.

To validate the use of parametric analysis, we first assessed whether the distribution of score differences satisfied the normality assumption. A Shapiro–Wilk test indicated no significant deviation from normality ($W = 0.95, p = 0.64$), supporting the appropriateness of subsequent parametric tests. We then examined the magnitude of learning gains using Cohen’s $d$, which yielded a large effect size ($d = 1.49$). This suggests that the observed improvement is not only statistically reliable but also practically meaningful, reflecting a substantial increase in subject matter competency.

\subsection{Statistical Correlations}
To compare baseline variability on database literacy, we surveyed students’ initial self-reported proficiency across key DBMS domains including SQL design, database types, ER modeling, and system use (see \cref{fig:DBLit}). 
Pearson correlations revealed no meaningful associations: gains were uncorrelated with AI literacy ($r=0.05$, $p=.88$) and weakly, non-significantly related to database literacy ($r=-0.09$, $p=.77$). These results indicate that initial confidence in AI or database skills did not predict who benefited most from the course, further suggesting that AI integration benefited all students irrespective of prior perceptions.

\subsection{Metacognitive Reflection and Engagement} Qualitatively, students' interactions with generative AI produced  evidence through their revisions and justifications of AI-generated artifacts. Course activities required students to inspect AI outputs, identify errors, and revise designs using database principles, reflecting metacognitive monitoring as students evaluated AI suggestions against task goals and domain knowledge rather than accepting them at face value.

Weekly interactions measured by page views averaged 63.9 (SD = 33.1; median = 63; range = 29--128), indicating variable engagement. Student feedback was largely positive: ``\textit{I really appreciated the way you ran the course, truly enjoyed and learnt here without just passing}'' and ``\textit{I have had a lot of fun with this class. Thank you so much!}'', suggesting the AI-supported format was engaging, though challenges with complex concepts, LLM hallucinations, and instruction-following raise the need for stronger scaffolding in both subject-matter and AI literacy.

\section{Threats to Validity}

We acknowledge several limitations that constrain the interpretation of our findings and outline how they may be addressed in future work.

\paragraph{Control Group.}
The absence of a control group and the small cohort size limit causal inference and statistical power. Accordingly, this study is positioned as a design-based investigation focused on instructional feasibility, coherence, and directional learning effects rather than comparative efficacy. While a randomized control condition would be required to isolate the incremental impact of AI scaffolds, the convergence of objective learning gains, self-reported literacy measures, and consistent improvements provides meaningful evidence of instructional impact aligned with established active-learning practices in CS education.

\paragraph{Sample Size and Statistical Power.}
The limited number of participants constrains our ability to detect small effects and increases the risk of Type II error. This limitation reflects course enrollment and the single-section deployment. Future studies should replicate the design across multiple sections or institutions to increase power. Nonetheless, the observed large pre/post learning gains indicate robust improvements well above the noise floor, suggesting pedagogically meaningful effects.

\paragraph{Self-Report and Calibration Bias.}
Measures of AI and database literacy rely on self-report and may be subject to over- or under-confidence, as reflected in the observed misalignment between perceived and objective competence. While future work could triangulate self-reports with behavioral or log-based measures, this dissociation itself is informative: it highlights the importance of calibration and reflective judgment when integrating AI into instruction, reinforcing the pedagogical relevance of our findings.

\section{Concluding Remarks}

This paper reframes generative AI as a pedagogical companion in undergraduate education, showing that structured use---via prompting, critique, and error analysis---can improve competency and metacognitive calibration. The framework generalizes beyond database design and motivates future controlled, multi-institution studies and adaptive AI scaffolds.

This study was constrained by methodological and sampling limitations.
The absence of a control group and the small sample size limit causal claims and statistical power, increasing the risk of Type II error. In addition, self-reported measures of AI and database literacy may be affected by calibration bias, as suggested by discrepancies between perceived and objective competence. Future work should extend this approach through larger multi-institution studies, controlled comparisons of AI-augmented and non-AI instruction, and adaptive AI scaffolding that responds to learner progression.

\subsection*{Acknowledgment}
This research is supported in part by NSF Award IIS-2144413. We thank the anonymous reviewers for their constructive comments.

\bibliographystyle{splncs04}
\bibliography{teachingAI}

\appendix

\section*{Supplementary Material}

\section{Evaluating AI Literacy and Competency} \label{sec:literacy}

We assessed students' AI competency using a two‑fold questionnaire. First, participants rated their own AI skills on a 5-point Likert scale (see \cref{tab:ai_literacy_questions}) resulting in a \textbf{self‑reported AI literacy score}. 
Second, they answered a series of questions on common AI topics such as prompt engineering, model limitations (e.g., hallucinations), and ethical considerations, which produced an \textbf{AI competency score} independent of personal perception (see \cref{tab:ai_competency_questions}). 
This dual approach enables us to capture learners' confidence in working with AI and to validate that confidence against demonstrable AI knowledge.

\subsection{AI Literacy}

We selected a subset of six questions to capture key facets of AI literacy relevant to our learning context (\cref{tab:ai_literacy_questions}). Two questions focused on \textbf{self-efficacy}, assessing students' confidence in using AI to solve complex tasks and achieve personal goals. Two additional questions targeted the \textbf{application of AI}, evaluating students' perceived ability to use AI tools in practical problem-solving and conceptual analysis. The remaining two questions addressed \textbf{ethical} awareness and the \textbf{evaluation} of AI, probing students' ability to assess its benefits and limitations. 
This selection provides a concise yet representative measure of students' AI literacy across cognitive, behavioral, and ethical dimensions.

\begin{table}[h!]
\scriptsize
\setlength{\tabcolsep}{5pt}
\centering
\caption{Self-report AI literacy questions}
\label{tab:ai_literacy_questions}
\begin{tabular}{p{0.79\linewidth} p{0.2\linewidth}}
\toprule
\textbf{Survey Question} & \textbf{Dimension} \\
\midrule
Q1. I can usually solve strenuous and complicated tasks when working with artificial intelligence well. & Self-efficacy \\ 
Q2. I think that users are responsible for considering AI design and decision processes. & Ethics \\
Q3. I can evaluate AI applications and concepts for different situations. & Application \\
Q4. I can use artificial intelligence meaningfully to achieve my goals. & Self-efficacy \\
Q5. I can apply AI applications to solve problems. & Application \\
Q6. I can assess what advantages and disadvantages the use of an artificial intelligence entails. & Evaluation \\
\bottomrule
\end{tabular}
\end{table}

\subsection{AI Competency}
We designed five multiple‑choice questions to objectively assess three core generative AI domains (\cref{tab:ai_competency_questions}): \textbf{conceptual knowledge} (understanding `tokens' and `prompts'), \textbf{prompt engineering} (few‑shot prompting and format instruction), and \textbf{model adaptation} (fine‑tuning pretrained models on task‑specific data). This concise set provides a clear snapshot of students' foundational, procedural, and applied AI competencies.

\begin{table}[h!]
\centering
\scriptsize
\caption{AI competency questions (choices omitted due to space)}
\label{tab:ai_competency_questions}
\begin{tabular}{p{0.95\linewidth}}
\toprule
\textbf{AI Multiple Choice Question} \\
\midrule
Q1. In the context of language models, what is meant by a ``token''? \\

Q2. In working with a large language model (LLM), what does the term ``prompt'' refer to? \\

Q3. What is meant by ``few-shot'' prompting in the context of generative AI? \\

Q4. What strategy is most effective for ensuring that an LLM produces output in a strict format (e.g., JSON)? \\

Q5. What does ``fine-tuning'' a pre-trained AI model involve? \\
\bottomrule
\end{tabular}
\end{table}

\section{Evaluating Subject Matter Literacy and Competency}\label{app:subjectmatter}

\subsection{Self-Reported DBMS Literacy}

We considered four questions to survey students' self‑reported knowledge of DBMS use, SQL design, database types, and ER schema (\cref{tab:db_selfreport_topics})

\begin{table}[h!]
\centering
\setlength{\tabcolsep}{5pt}
\scriptsize
\caption{Self‑report DBMS questions}
\label{tab:db_selfreport_topics}
\begin{tabular}{p{0.75\linewidth}l}
\toprule
\textbf{Survey Question}                                        & \textbf{Topic}  \\
\midrule
Q1. I know how to use a database management system.                              & DBMS use        \\
Q2. I can design and manipulate a database via SQL effectively.                  & SQL design      \\
Q3. I am familiar with different types of databases (relational, document, graph). & DB types        \\
Q4. I can design a relational database schema using ER diagrams.                 & ER schema       \\
\bottomrule
\end{tabular}
\end{table}

\subsection{Subject Matter Competency}

Seven core questions, spanning conceptual understanding, SQL query formulation, and relational vs. NoSQL reasoning, were first administered as an ungraded, low‑stakes pre‑test. 
The identical or minimally revised questions reappeared across the midterm and final exams, enabling direct comparison of each student's performance before and after instruction.

\section{Additional Quantitative Analysis}\label{sec:quant}

\subsection{Analyzing AI Literacy and Competency}
We first analyze students agreement/disagreement on AI literacy questions. 
\textit{Application} (Q3 \& Q5) sits at 50\% average agreement; roughly half remain neutral, highlighting an opportunity for more hands‑on, context‑driven AI problem‑solving tasks.
\textit{Self‑Efficacy} (Q1 \& Q4) is the weakest dimension (45.5\% average agreement and 4\% average disagreement), revealing that many students are uncertain about their ability to tackle complex AI tasks or use AI meaningfully—pointing to the need for scaffolded labs and goal‑oriented prompt‑engineering workshops.
\textit{Ethics} (Q2) shows the strongest consensus: 67\% agreement and no disagreement, indicating clear student buy‑in on their responsibility to consider AI’s design and decision processes.
\textit{Evaluation} (Q6) is with 66\% agreement on assessing AI’s advantages and disadvantages, though 8\% disagree, suggesting targeted critique exercises could further shore up this skill.
The descriptive statistics of AI competency scores for questions shown in \cref{tab:ai_competency_questions} are clustered in domains and illustrated in \cref{tab:ai_competency_domains}.

\begin{table}[t]
\centering
\scriptsize
\caption{AI competency score by domain}
\label{tab:ai_competency_domains}
\begin{tabular}{lccc}
\toprule
\textbf{Domain} & \textbf{Q} & \textbf{Mean \%} & \textbf{Std. Dev.} \\
\midrule
Conceptual Knowledge & Q1, Q2 & 75.0 & 24.04 \\
Prompt Engineering   & Q3, Q4 & 62.5 & 28.99 \\
Model Adaptation     & Q5     & 75.0 & 0.00 \\
\bottomrule
\end{tabular}
\end{table}

\subsection{Pre/Post Assessment by Question}
Table \ref{tab:prepost_per_question} shows that in almost all questions the performance has significant improvement ($p<0.05$), with large t-statistics and substantial mean gains (ranging $0.46-0.77$). The only exceptions were Q1 (DBMS use) and Q4 (basic normalization) due to high score in pre-assessment.

\begin{table}[h!]
\centering
\setlength{\tabcolsep}{10pt}
\scriptsize
\caption{Pre/Post assessment statistics, divided by questions.}
\label{tab:prepost_per_question}
\begin{tabular}{l c c c c c}
\toprule
\textbf{Question} & \textbf{Pre} & \textbf{Post} & \textbf{$t$} & \textbf{$p$ value}\\
& Mean (SD) & Mean (SD) & \textbf{stat} & \\
\midrule
Q1 & 0.833 (0.15) & 0.917 (0.083) & 0.561 & 0.586 \\
Q2 & 0.333 (0.24) & 0.917 (0.083) & 3.023 & 0.0116 \\
Q3 & 0.417 (0.26) & 0.917 (0.083) & 2.569 & 0.0261 \\
Q4 & 0.750 (0.20) & 1.000 (0.00) & 1.915 & 0.0819 \\
Q5 & 0.667 (0.24) & 1.000 (0.00) & 2.345 & 0.0388 \\
Q6 & 0.583 (0.26) & 1.000 (0.00) & 2.803 & 0.0172 \\
Q7 & 0.667 (0.24) & 1.000 (0.00) & 2.345 & 0.0388 \\
\bottomrule
\end{tabular}
\end{table}

\end{document}